\documentclass[twocolumn,showpacs,amsfonts,aps,prc,nofootinbib,floatfix,%
superscriptaddress]{revtex4}

\usepackage{amsmath}
\usepackage{bm}
\usepackage{graphicx}

\usepackage{epsfig}
\newcommand{\beq}{\begin{equation}}
\newcommand{\eeq}{\end{equation}}
\newcommand{\bea}{\vspace{0.25cm}\begin{eqnarray}}
\newcommand{\eea}{\end{eqnarray}}

\newcommand{\ro}{\mbox{{\boldmath
$\rho$}}}
\newcommand{\rr}{\mbox{{\boldmath
$\rho$}}}
\newcommand{\Pb}{\mbox{{\bf
P}}}

\newcommand{\pb}{\mbox{{\bf
p}}}

\newcommand{\kb}{\mbox{{\bf
k}}}

\newcommand{\qbt}{\mbox{{\bf
q}}_\perp}

\newcommand{\Fb}{\mbox{{\bf
F}}}

\newcommand{\fb}{\mbox{{\bf
f}}}

\def\lsim{\mathrel{\rlap{\lower4pt\hbox{\hskip1pt$\sim$}}
    \raise1pt\hbox{$<$}}}         
\def\gsim{\mathrel{\rlap{\lower4pt\hbox{\hskip1pt$\sim$}}
    \raise1pt\hbox{$>$}}}         

\newcommand{\landau}{L.D.~Landau Institute for Theoretical Physics,
        GSP-1, 117940, Kosygina Str. 2, 117334 Moscow, Russia}

\begin{document}


\title{
Synchrotron contribution to photon emission
from quark-gluon plasma
}
\date{\today}

\author{B.G.~Zakharov}\affiliation{\landau}

\begin{abstract}
We study the influence of the magnetic field on the photon emission
from the quark-gluon plasma created in $AA$ collisions.
We find that even for very optimistic assumption
on the magnitude of the magnetic field for noncentral $AA$ collisions
the effect of magnetic field is very small.
\end{abstract}
%

\maketitle

\section{Introduction}
Experimental study of photon  
spectra in the low and intermediate $k_T$ region in $AA$ collisions
can provide vital information on the parameters of the produced
quark-gluon plasma (QGP) \cite{Shuryak}.
It is widely believed
that the observed in $AA$ collisions at RHIC 
\cite{PHENIX_ph1,PHENIX_ph_v2,PHENIX_ph_PR} 
and LHC \cite{ALICE_ph} excess of the photon
yield (above the photons from hadron decays and from the hard perturbative
mechanism)  at $k_T\lsim 3-4$ GeV is related to photon emission
from the QGP. It is surprising that the thermal photons exhibit a significant 
azimuthal asymmetry $v_2$ (``elliptic flow'')
comparable to that for hadrons.
It is difficult to reconcile with the expectation that the thermal photons
should be mostly radiated from the hottest initial stage of the QGP
where the flow effects should be small (it is usually called
``the direct photon puzzle'').

It was suggested by Tuchin \cite{T1} that the azimuthal
anisotropy of the direct photons may be due to 
synchrotron mechanism of the photon emission
in a strong transverse (to the reaction plane)
magnetic field in the noncentral $AA$ collisions.
The synchrotron contribution obtained in \cite{T1} can explain a 
significant fraction of the photon yield
in the central rapidity region at $k_T\sim 1-3$ GeV.
However, the calculations performed in \cite{T1} 
are of a qualitative nature.
In particular in \cite{T1} it was ignored the fact that multiple
scattering of quarks, which they undergo in the thermal bath, 
will suppress the synchrotron emission (because of 
reduction of the coherence/formation length of the photon
emission). 
In fact for the photon emission in the QGP with magnetic field
one cannot distinguish between the synchrotron radiation and
the bremsstrahlung due to multiple scattering. One can only define 
the difference between the photon emission rate from the QGP 
with and without magnetic field. 
On the other hand, in \cite{T1} there was not taken into account
the contribution of the synchrotron annihilation $q\bar{q}\to \gamma$
which increases the the photon emission.
It is known that for the QGP without magnetic field
the annihilation contribution is more important than bremsstrahlung 
at the photon momenta $k_T\gg T$ \cite{AMY1}. 

In the present work we address the effect of the magnetic
field on both the processes $q\to \gamma q$ and $q\bar{q}\to \gamma$.
We develop a formalism which treats on an even footing 
the effect of multiple scattering and curvature of the quark trajectories
in the collective magnetic field in the QGP. 
Our analysis is based on the light cone
path integral (LCPI) formalism \cite{LCPI}, which was previously
successfully used \cite{AZ} for very simple derivation
of the well known photon emission rate from the higher order collinear processes
$q\to \gamma q$ and $q\bar{q}\to \gamma$ obtained by Arnold, Moore and Yaffe
(AMY) \cite{AMY1} using methods from thermal field theory 
with Hard Thermal Loop (HTL) resummation.
It is known that the higher order diagrams 
corresponding to these processes contribute to leading order \cite{AGZ2000},
and turn out to be as important as the LO  
$2\to 2$ processes $q(\bar{q})g\to \gamma q(\bar{q})$ (Compton)
and $q\bar{q}\to \gamma g$ (annihilation) \cite{Baier_ph}. 
Contrary to the collinear processes the LO processes 
should not be affected by the presence of the magnetic field.
Our results differ drastically from that of 
\cite{T1}. We find that even for very optimistic
magnitude of the magnetic field for RHIC and LHC conditions
the effect of the magnetic field on the photon
emission from the QGP is very small.

\section{The processes $q\to \gamma q$ and $q\bar{q}\to \gamma$
in the QGP with magnetic field}

As in \cite{AZ} 
we treat quarks as a relativistic particles 
with $p\gg m_q$, where $m_q$ is the thermal quark quasiparticle 
mass. 
The same approximation is used in the AMY analysis \cite{AMY1}.
For relativistic quarks, similarly to the QGP without magnetic field, 
the processes $q\to \gamma q$ and $q\bar{q}\to \gamma$ are dominated 
by the collinear configurations, when the photon is emitted practically 
in the direction of the initial quark for $q\to \gamma q$ (and
in the direction of the momentum of the $q\bar{q}$ pair
for $q\bar{q}\to \gamma$).
The contribution of the collinear processes to the photon emission 
rate per unit time and volume can be written as \cite{AZ,AMY1}
\beq
\frac{dN}{dtdVd\kb}=
\frac{dN_{br}}{dtdVd\kb}
+\frac{dN_{an}}{dtdVd\kb}\,,
\label{eq:10}
\eeq
where the two terms correspond to $q\to \gamma q$ and $q\bar{q}\to \gamma$
processes.
The contribution of the bremsstrahlung mechanism reads \cite{AZ}
\bea
\frac{dN_{br}}{dtdVd\kb}=\frac{d_{br}}{k^{2}(2\pi)^{3}}
\sum_{s}
\int_{0}^{\infty} dp p^{2}n_{F}(p)\nonumber\\
\times
[1-n_{F}(p-k)]\theta(p-k)
\frac{dP^{s}_{q\rightarrow \gamma q}(\pb,k)}{dk dL}\,,
\label{eq:20}
\eea
where 
$d_{br}=4N_{c}$ is the number of the quark and antiquark states,
$n_{F}(p)=1/[\exp(p/T)+1]$
is the thermal Fermi distribution, 
and 
${dP^{s}_{q\rightarrow \gamma q}(\pb,k)}/{dk dL}$
is the probability  of the photon emission 
per unit length from a fast quark of type $s$ interacting with the 
random soft gluon field
generated by the thermal partons and with the external smooth electromagnetic 
field (which generates the Lorentz force $\Fb$).
In the small angle approximation the vectors $\pb$  and $\kb$ are parallel.
So the problem is reduced to calculation of 
$dP^{s}_{q\rightarrow \gamma q}(\pb,k)/{dk dL}$.
The annihilation contribution is related to the photon 
absorption via the detailed balance principle 
as \cite{AZ}
\beq
\frac{dN_{an}}{dtdVd\kb}=
[1+n_{B}(k)]^{-1}
\frac{dN_{abs}}{dtdVd\kb}\,.
\label{eq:30}
\eeq
The photon absorption rate on the right-hand side of (\ref{eq:30})
can be written via the probability distribution per unit length
for the $\gamma \rightarrow q\bar{q}$ transition
$
{dP^{s}_{\gamma\rightarrow q\bar{q}}(\kb,p)}/{dp dL}
$, where $p$ is the
final quark momentum.
Then one
obtains \cite{AZ}
\bea
\frac{dN_{an}}{dtdVd\kb}=\frac{d_{an} }{(2\pi)^{3}}
\sum_{s}
\int_{0}^{\infty} dp n_{F}(p)\nonumber\\
\times n_{F}(k-p)\theta(k-p)
\frac{dP^{s}_{\gamma\rightarrow q\bar{q}}(\kb,p)}{dp dL}\,,
\label{eq:40}
\eea
where $k-p$ is the antiquark momentum, $d_{an}=2$ is the number
of the photon helicities.

Let us consider first calculation of the bremsstrahlung contribution. 
In the LCPI formalism \cite{LCPI} 
the probability of the $q\to \gamma q$ transition
(for a quark with charge $z_{q}e$) per unit length can be written
in the form
(we use here the fractional photon momentum $x$ instead of 
$k$)
\bea
\frac{d P_{q\rightarrow \gamma q}^{s}}{d
x dL}=2\mbox{Re}
\int\limits_{0}^{\infty} d
z
\exp{\left(-i\frac{z}{L_{f}}\right)}\nonumber\\
\times \left.\hat{g}(x)\left[
{\cal K}(\ro_{2},z|\ro_{1},0)
-{\cal K}_{vac}(\ro_{2},z|\ro_{1},0)
\right]\right|_{\ro_{1,2}=0}\,,
\label{eq:50}
\eea
where $L_{f}=2M(x)/\epsilon^{2}$ with $M(x)=E_qx(1-x)$,
$\epsilon^{2}=m_{q}^{2}x^{2}+m_{\gamma}^{2}(1-x)$
(in general for $a\to b+c$ transition 
$\epsilon^{2}=m_{b}^{2}x_{c}+m_{c}^{2}x_{b}-m_{a}^{2}x_{b}x_{c}$),
$\hat{g}$ is the vertex operator, given by
\beq
\hat{g}(x)=\frac{g_{1}(x)}{M^{2}(x)}\frac{\partial }{\partial \ro_{1}}\cdot
\frac{\partial }{\partial \ro_{2}}\,
\label{eq:60}
\eeq
with 
\beq
g_{1}(x)=z_{q}^{2}\alpha_{em}(1-x+x^{2}/2)/x. 
\label{eq:70}
\eeq
$\cal{K}$ in (\ref{eq:50}) is the Green function for the Hamiltonian
\beq
\hat{\cal{H}}=-\frac{1}{2M(x)}
\left(\frac{\partial}{\partial \ro}\right)^{2}
+         v(\ro)\,,
\label{eq:80}
\eeq
and ${\cal{K}}_{vac}$ is the Green function for $v=0$.
The potential reads
\beq
v=v_{f}+v_{m}\,,
\label{eq:90}
\eeq
where $v_{f}$ is due to the fluctuating gluon fields of the QGP,
and $v_{m}$ is related to the mean electromagnetic field.
The mean field component of the potential reads
\beq
v_{m}=-\fb \ro\,,
\label{eq:100}
\eeq
where $\fb=x z_{q}\Fb$, $\Fb$ is transverse component (to the parton momentum) 
of the Lorentz force for a particle with unit charge.
The effect of the longitudinal Lorentz force (which exists for non-zero
electric field) is small for the relativistic partons,
and we neglect it.
The component $v_f$ reads  
\beq
v_{f}=-i P(x\rho)\,.
\label{eq:110}
\eeq
Here the function $P(\rho)$ 
can be written as
\beq
P(\ro)=g^{2}C_{F}\int\limits_{-\infty}^{\infty} dz 
[G(z,0_{\perp}z)-G(z,\ro,z)]\,,
\label{eq:120}
\eeq
where $g$ is the QCD coupling, $C_F=4/3$ is the quark Casimir,
the gluon correlator $G$ (the color indexes are omitted) reads
\beq
G(x-y)= 
u_{\mu}u_{\nu}
{\Large\langle\Large\langle}
A^{\mu}(x)A^{\nu}(y)
{\Large\rangle\Large\rangle}\,
\label{eq:130}
\eeq
where $u^{\mu}=(1,0_{\perp},1)$ is the light-like vector along
the $z$ axis (we define the $z$ axis along the initial quark momentum).
In the HTL scheme one can obtain \cite{PA_C}
\beq
P(\ro)= \frac{g^{2}C_{F}T}{(2\pi)^{2}}\int d\qbt [1-\exp(i\ro \qbt)]
C(\qbt)\,,
\label{eq:140}
\eeq
\beq
C(\qbt)=\frac{m_{D}^{2}}{\qbt^{2}(\qbt^{2}+m_{D}^{2})}\,,
\label{eq:150}
\eeq  
where $m_{D}=gT[(N_{c}+N_{F}/2)/3]^{1/2}$ 
is the Debye mass.
In the approximation (in the sense of multiple scattering in the QGP)
of static color Debye-screened scattering centers the function $P(\ro)$ reads
\beq
P(\ro)= 
\frac{n{\sigma}_{q\bar{q}}(\rho )}{2}\,,
\label{eq:160}
\eeq
where $n$ is the number density
of the color centers, and 
\beq
\sigma_{q\bar{q}}(\rho)={C_{T}C_{F}\alpha_{s}^{2}}\int d\qbt
\frac{[1-\exp(i\qbt\ro)]}{(\qbt^{2}+m_{D}^{2})^{2}}\,\,
\label{eq:170}
\eeq
is the well known dipole cross section \cite{NZ12}
with $C_T$ being the color center Casimir.

Both for the HTL scheme and the static approximation at $\rho\lsim 1/m_D$ 
approximately $P(\rho)\propto \rho^2$.
We will work in the oscillator approximation 
\beq
P(\rho)=C_{p}\rho^{2}\,.
\label{eq:180}
\eeq
The $C_{p}$ can be expressed via  the well known transport 
coefficient $\hat{q}$ \cite{BDMPS}
$C_p=\hat{q}C_{F}/4C_{A}$. 
Qualitative pQCD calculations give $\hat{q}\sim 2\varepsilon^{3/4}$ 
\cite{Baier_qhat}, 
where $\varepsilon$ is the QGP energy density. It gives
$\hat{q}\approx 0.2$ GeV$^{3}$ 
at $T=250$ MeV. This agrees well with estimate
of $\hat{q}$ via the form of the dipole cross section at small
$\rho$ that allows to describe well the data on jet quenching 
in $AA$ collisions
within the LCPI scheme \cite{RAA11,RAA12,RAA13}.

For the quadratic $P(\rho)$ the  the Hamiltonian (\ref{eq:80})
takes the oscillator form
\beq
\hat{\cal{H}}=-\frac{1}{2M(x)}
\left(\frac{\partial}{\partial \ro}\right)^{2}
+   \frac{M\Omega^{2}\ro^2}{2}-\fb\ro\,
\label{eq:190}
\eeq
with
\beq
\Omega=\sqrt{-iC_{p}x^{2}/M}\,.
\label{eq:200}
\eeq
The Green function for the Hamiltonian (\ref{eq:190}) is known explicitly 
(see, for example, \cite{FH})
\beq
{\cal{K}} (\rr_2 , z_2 | \rr_1 , z_1)=
\frac{M\Omega}{2\pi i\sin(\Omega \Delta z)}
\exp{[i S_{cl}]}\,,
\label{eq:210}
\eeq
where $\Delta z=z_2-z_1$ and $S_{cl}$ is the classical action. 
The action can be
written as a sum  $S_{cl}=S_{osc}+S_{f}$ with
\beq
\hspace{-.05cm}S_{osc}=
\frac{M\Omega}{2\sin(\Omega \Delta z)}
\left[\cos(\Omega \Delta z)(\rr_{1}^{2}+\rr_{2}^{2})-2\ro_{1}\ro_{2}\right]\,,
\label{eq:220}
\eeq
\beq
S_{f}
=
\frac{M\Omega}{2\sin(\Omega \Delta z)}
\left[\Pb(\ro_{1}
+\ro_{2})-W\right]\,,
\label{eq:230}
\eeq
where
\beq
\Pb=\frac{2\fb}{M\Omega^{2}}
[1-\cos(\Omega \Delta z)]\,\,,
\label{eq:240}
\eeq
\beq
W=\frac{2\fb^{2}}{M^{2}\Omega^{4}}
\left[1-\cos(\Omega \Delta z)-\frac{\Omega \Delta z
\sin(\Omega \Delta z)}{2}\right]\,.
\label{eq:250}
\eeq

Then, after including the vacuum term  a simple calculation gives
\beq
\frac{dP}{dxdL}=2g_1 (I_{osc}+I_{s})\,.
\label{eq:260}
\eeq
Here $I_{osc}$ corresponds to the pure oscillator case
($\fb=0$). It reads
\bea
I_{osc}=\frac{1}{\pi}\mbox{Re}\int_{0}^{\infty} dz
\left[
\frac{1}{z^2}-\left(\frac{\Omega}{\sin(\Omega z)}\right)^{2}
\right]\nonumber\\
\times\exp\left(-i\frac{z}{L_{f}}\right)\,.
\label{eq:270}
\eea
And $I_s$ gives the synchrotron correction.
It can be written as a sum $I_s=I_1+I_2$ with 
\beq
I_{1}=
\frac{1}{\pi}\mbox{Re}\int_{0}^{\infty} dz
\left(\frac{\Omega}{\sin(\Omega z)}\right)^{2}
[1-\exp(-U)]\,,
\label{eq:280}
\eeq
\bea
I_{2}=
\frac{1}{\pi}\mbox{Re}\int_{0}^{\infty} dz
\frac{iM\Omega^{3}}{8\sin^{3}(\Omega z)}
\Pb^2
\exp\left(-U-i\frac{z}{L_{f}}\right)\,,
\label{eq:290}
\eea
where 
$U={iM\Omega W}/{2\sin(\Omega z)}\,$ (here $z$ corresponds to $\Delta z$
in (\ref{eq:210})-(\ref{eq:250})).
In the limit $\Omega\to0$ $I_{osc}$ vanishes. In this limit
$I_{s}$ can be expressed via the Airy function, and the radiation rate
is reduced to the well known quasiclassical formula
for the synchrotron spectrum \cite{BK}.

For $\gamma\to q \bar{q}$ one can obtain similar formulas.
But now $M(x)=E_{\gamma}x(1-x)$ ($x$ is the quark fractional momentum)
$\epsilon^{2}=m_{q}^{2}-m_{\gamma}^{2}x(1-x)$,
$\fb=z_{q}\Fb$, and
\beq
g_{1}=z_{q}^{2}\alpha_{em}[x^{2}+(1-x)^{2}]/2\,,
\label{eq:300}
\eeq
\beq
\Omega=\sqrt{-iC_{p}/M}\,.
\label{eq:310}
\eeq

We perform calculations for standard quark and
photon quasiparticle masses in the QGP
$m_q=gT/\sqrt{3}$ and $m_{\gamma}=\frac{eT}{3}\sqrt{(3+N_f)/2}$
\cite{AMY1}. We take $N_f=2.5$ to account for qualitatively the mass
suppression for strange quarks at moderate temperatures.

\section{Numerical results}

\begin{figure} [h]
\begin{center}
\epsfig{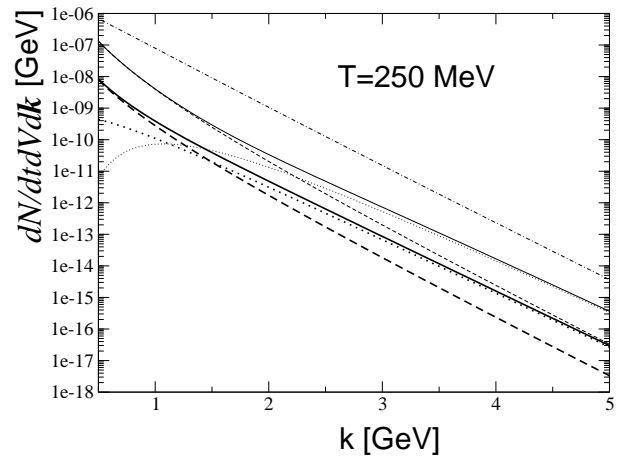}
\end{center}
\caption[.]{The synchrotron contribution to the photon emission rate
$dN/dtdVd\kb$ 
from $q\to \gamma q$ (dashed) and $q\bar{q}\to \gamma$ (dotted)
processes and their sum (solid)
at $T=250$ MeV and $eB=m_{\pi}^2$ obtained with (thick lines) and without
(thin lines) the effect of multiple scattering. The dash-dotted line shows the
contribution from the LO  $2\to 2$ processes.}
\end{figure}
\begin{figure} [h]
\begin{center}
\epsfig{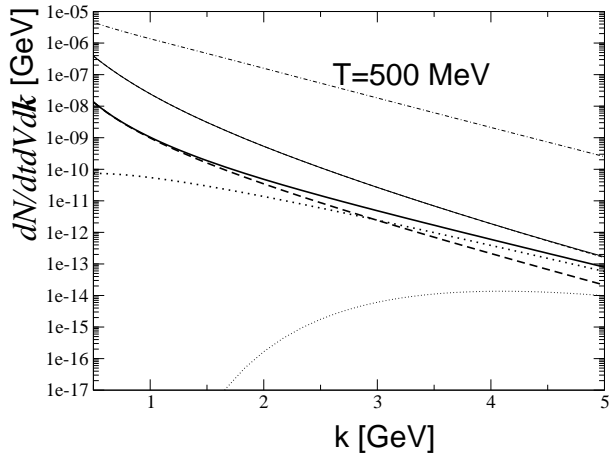}
\end{center}
\caption[.]{The synchrotron contribution to the photon emission rate
$dN/dtdVd\kb$ 
from $q\to \gamma q$ (dashed) and $q\bar{q}\to \gamma$ (dotted)
processes and their sum (solid)
at $T=500$ MeV and $eB=m_{\pi}^2$ obtained with (thick lines) and without
(thin lines) the effect of multiple scattering. The dash-dotted line shows the
contribution from the LO  $2\to 2$ processes.}
\end{figure}

The magnetic field in the noncentral $AA$ collisions is mostly
perpendicular to the reaction plane (this direction
corresponds to $y$ axis, if $x$ axis is directed along the impact parameter 
of $AA$ collisions). For this reason the transverse
(to the quark momentum) component of the Lorentz force is 
$\propto \cos(\theta)$. This fact leads naturally to a strong
azimuthal asymmetry $v_2$ for the synchrotron radiation \cite{T1}. 
This effect can be only observed if the relative contribution of 
the synchrotron mechanism to the photon emission rate is not very
small.

For Au+Au collisions at $\sqrt{s}=0.2$ TeV
the typical value of the magnetic field at $b\sim 6$ fm and proper 
time $\tau\sim 0.2$ fm is $eB\sim 0.1 m_{\pi}^2$ \cite{Z_B}
\footnote{Note that for Pb+Pb collisions at $\sqrt{s}=2.76$ TeV
the magnetic field is stronger only at very low values of $\tau$,
that are of no interest from the point of view of the photon emission from
the QGP. For $\tau\gsim 0.1$ fm which may be of interest to us
the magnetic field is smaller than for Au+Au collisions 
at $\sqrt{s}=0.2$ TeV \cite{Z_B}.}.
We perform numerical calculations for more optimistic value 
$eB=m_{\pi}^2$. In Figs.~1,~2 we present the results for the effect
of the magnetic field on the photon emission rate
for $T=250$ and $500$ MeV. We show the results separately
for bremsstrahlung and annihilation and for their sum.
We present also the curves obtained neglecting
the effect of multiple scattering ($\Omega=0$).
For the comparison we present in Figs.~1,~2 the contribution 
of the LO mechanisms in the form obtained in \cite{AMY1}.
One sees that multiple scattering suppresses strongly the contribution
of the synchrotron radiation. 
The curves for the synchrotron mechanism go considerably below the
ones for the LO contribution. And for a version with multiple scattering
the contribution of the synchrotron mechanism turns out to be practically
negligible as compared to the LO mechanism. 
We see that even for our clearly too optimistic value of the magnetic
field the effect of the synchrotron mechanism is 
very small. For more realistic field $eB\sim 0.1 m_{\pi}^2$
the synchrotron contribution is smaller by a factor of $\sim 10^2$.
Thus, one can conclude that the effect of the magnetic field
cannot be important for photon emission in $AA$ collisions.

\section{Summary}
We have studied the influence of the magnetic field
on the photon emission rate from the QGP.
We find that even for clearly too optimistic 
assumption on the magnitude of the magnetic field
($eB\sim m_{\pi}^2$)
the effect of magnetic field
is very small, and for more realistic
fields ($eB\sim 0.1 m_{\pi}^2$) the effect is practically negligible.
For this reason we conclude that 
the synchrotron mechanism cannot solve 
``the direct photon puzzle''.
Thus, our calculations do not support the results of the recent analysis
\cite{T1}, where a rather large effect of magnetic
field was found.

\begin{acknowledgments} 	
I thank P. Aurenche for useful discussions in the initial stage
of this work.
This work is supported by the Russian Scientific Foundation (grant 
No. 16-12-10151).
\end{acknowledgments}

\end{document}